\documentclass[aps,pre,twocolumn,superscriptaddress]{revtex4-2}
\usepackage{amssymb}
\usepackage{amsmath,bm}
\usepackage{graphicx,color}
\usepackage{bbm}
\usepackage[bottom]{footmisc}
\usepackage{multirow}
\usepackage{xcolor}
\usepackage{xpatch} 
\makeatletter   
\usepackage{xpatch} 

\newcommand{\B}[1]{{\bm{#1}}}

\newcommand{\Lag}{\mathcal{L}}
\newcommand{\A}{\mathcal{A}}
\newcommand{\dif}{\mathrm{d}}
\newcommand{\rin}{r_\text{in}}
\newcommand{\rout}{r_\text{out}}
\newcommand{\C}[1]{{\mathcal{#1}}}

\begin{document}
\title{Anomalous Elasticity and Emergent Dipole Screening in Three-Dimensional Amorphous Solids}
\author{Harish Charan}
\affiliation{Dept. of Chemical Physics, The Weizmann Institute of Science, Rehovot 76100, Israel}
\author{Michael Moshe}
\affiliation{Racah Institute of Physics, The Hebrew University of Jerusalem, Jerusalem, Israel 9190}
\author{Itamar Procaccia} 
\affiliation{Dept. of Chemical Physics, The Weizmann Institute of Science, Rehovot 76100, Israel}
\affiliation{Center for OPTical IMagery Analysis and Learning, Northwestern Polytechnical University, Xi'an, 710072 China}

\begin{abstract}
In recent work, we developed a screening theory for describing the effect of plastic events in amorphous solids on its emergent mechanics. The suggested theory uncovered an anomalous mechanical response of amorphous solids where plastic events collectively induce distributed dipoles that are analogous to dislocations in crystalline solids. The theory was tested against various models of amorphous solids in two-dimensions, including frictional and friction-less granular media and numerical models of amorphous glass.
Here we extend our theory to screening in three-dimensional amorphous solids and predict the existence of anomalous mechanics similar to the one observed in two-dimensional systems. 
We conclude by interpreting the mechanical response as the formation of non-topological distributed dipoles that have no analogue in the crystalline defects literature. Having in mind that the onset of dipole screening is reminiscent of Kosterlitz-Thouless and Hexatic transitions, the finding of dipole screening in three-dimensions is particularly novel. 
\end{abstract}
\maketitle
\section{Introduction}

Amorphous solids include materials like glasses and granular media. These materials lack long-range order, and they do not possess a unique ``ground state". They can be cooled down to zero temperature, where they can reside in one of many available local equilibria, which under mechanical strains can easily exchange relative stability. Research in the last decade or two indicated that classical elasticity theory needs to be reconsidered for the treatment of amorphous solids. Contrary to perfect crystalline solids that can exhibit purely elastic response to strain, amorphous solids suffer from plastic responses, and these appear as quadrupolar Eshelby-like structures \cite{54Esh,99ML,06ML}. It should be noted that this quadrupolar symmetry of localized plastic events stems from a purely geometric
conservation law, forbidding the formation of monopolar
and dipolar responses; quadrupolar plastic events are
the lowest order non-conserved multipoles  \cite{15KMS}. In amorphous solids plasticity appears (in the thermodynamic limit) at any infinitesimal deformation \cite{10KLPb,11HKLP}. Consequently nonlinear elasticity needs to be reconsidered as well; the higher order elastic moduli have sample-to-sample fluctuations which can diverge upon increasing the system size \cite{11HKLP,16PRS,16DPSS,16BU,17DIPS}. The upshot of these findings is that plastic events cannot be neglected in formulating a mechanical theory of amorphous solids. 

In a series of recent publications, it was discovered that the prevalence of plastic events in two-dimensional amorphous solids results in screening phenomena that are akin, but richer and different, to screening effects in electrostatics \cite{21LMMPRS,22MMPRSZ,22BMP,22KMPS}. Plastic events, which are localized quadrupole singularities in a reference curvature field \cite{15KMS, 15MLAKS},
and consequently typically quadrupoles in the displacement field, can act as screening charges. It was shown that when the density of plastic quadrupoles is low, their effect is limited to renormalizing the elastic moduli, but the structure of linear elasticity theory remains intact. This is analogous to dipole screening in dielectrics, where the dielectric constant gets renormalized by nucleation of localized dipole charges. It was shown that when the nucleation cost of quadrupoles
in an amorphous solid is finite, and consequently the density of plastic quadrupoles is low, their effect is limited to renormalizing the elastic moduli, with the structure of
linear elasticity theory remaining intact. But when the nucleation cost of quadrupoles drops, the quadrupoles density
becomes high, and the nucleation of effective dipoles defined by the gradients of their density, cannot be neglected. The presence of effective dipoles has surprising
consequences, changing the analytic form of the response
to strains in ways that cannot possibly be predicted by
standard elasticity theory. It was concluded that in two
dimensions one needs to consider a new theory, and this
emergent theory was confirmed by comparing its predictions to results of extensive experiments and simulations
\cite{21LMMPRS,22MMPRSZ,22BMP,22KMPS}.

We note that `screening' in solids has a fascinating connection with the melting transition in 2-dimensional crystals. Nelson and Halperin showed that a crystal melts via a sequential Kosterlitz-Thouless type phase transition with an intermediate Hexatic phase characterized by quasi-long-range orientational order and short-range positional order \cite{16Kos, 79NH}. From a pure mechanical perspective, the nucleation cost of a dislocation pairs that are quadrupoles of the reference curvature, is negligible in the Hexatic phase, and the nucleation cost for dislocations, which  are dipoles of the reference curvature, becomes finite, leading to a mechanical response that is screened by the emergent dipoles \cite{19ZR}. It is not yet clear whether or not the transition from quadrupole to dipole-screening in athermal amorphous solids is of the Kosterlitz-Thouless type. Regardless of the nature of the transition, the Hexatic phase is based on the existence of point dislocations in two-dimensional systems, and therefore irrelevant to three-dimensional solids. A question of major interest that was left open, is whether in three-dimensional amorphous solids one should expect dipole screening effects.

The aim of this paper is to answer this question in the affirmative, and to provide a theory of what we term ``anomalous elasticity", which in the context of three-dimensional amorphous solids takes into explicit account the existence of prevalent plastic responses. As in our work in two-dimensions, we will study systems in mechanical equilibria in which a source of strain is added, and the main question will be what is the resulting screened displacement field \cite{21BKSZ}.

To illustrate the emerging theory we will consider the response to a local stress increase in an amorphous solid made from a binary mixture of $N$ spherical balls of radii $r_a$ and $r_b$ respectively, 
which are contained in a spherical outer boundary of a large radius $r_{\rm out}$. At the origin there is an inner boundary of radius $r_{\rm in}$ which is of the order of the radii of the balls, $r_{\rm in}\ll r_{\rm out}$.  Details of interactions and simulation protocols for the system preparation are provided in Appendix~\ref{proto}. Once brought to mechanical equilibrium at a target pressure, the inner boundary is inflated by a chosen (small) percentage, and we examine the displacement field that is induced by this inflation. In a normal elastic material in spherical geometry we expect the radial component of the displacement field to be outward directed and to decay like $1/r^2$ where $r$ is the distance from the inner boundary. Contrary to classical elasticity, we observe here that the actual displacement fields exhibit much richer functional forms, including oscillations, and inward directed regions, to mention a few. We shall see that the radial component of the displacement field can even {\em increase} when
$r$ increases, in a striking contradiction with the classical elasticity expectation. The theory developed in this paper
explains fully and quantitatively this observed behavior and other
similar novel results.

The structure of this paper is as follows: In Sect.~\ref{elastic} we remind the reader of the standard theory of the spatial dependence of displacement fields induced by a local elastic charge.  The third section \ref{charges} geometric charges in 3-dimensions. Sect.~\ref{dilquad} will present screening by a dilute set of quadrupolar responses. We will see that dilute plastic responses can only renormalize the elastic moduli. Sect.~\ref{highdensity}
deals with the situation of higher densities of quadrupoles that induce the emergence of effective dipoles. This section will lead to the theory of anomalous
elasticity.  The last section \ref{summary} will offer a summary of the paper and indications for the road ahead.

\section{Displacement fields in classical elasticity}
\label{elastic}

The approach in this paper follows closely the two-dimensional counterpart as detailed in Ref.~\cite{21LMMPRS}. To derive the equations for the displacements field we identify an effective total energy that describes how the plastic Eshelby-like quadrupoles interact with the background elastic field and among themselves~\cite{12DHP,13DHP}. Minimizing this energy will provide the wanted equations.

We start in the next subsection with the classical formalism that accounts for the response of a purely elastic systems to strains on its boundaries.

\subsection{Minimizing the total energy}

In the context of classical elasticity \cite{Landau}, the energy $F$ is written as a sum of two contributions, an internal energy $U$ and the work done by traction forces $\B t$ on the boundary. $U$ is an integral over the stress times the strain. The strain and stress tensors are denoted $u_{\alpha\beta}$,  $\sigma^{\alpha\beta}$ and the displacement field by $d_\alpha$. The energy density is denoted as $\C L$. In three dimensions the total energy $F$ is 
\begin{eqnarray}
F &=& \int \Lag \, \mathrm{d}^3 x - \oint t^\beta  d_\beta \, \mathrm{d}S \ , \nonumber\\
\C L &=&  \frac{1}{2} A^{\alpha\beta\gamma\delta} u_{\alpha\beta}u_{\gamma\delta} =\frac{1}{2}\sigma^{\alpha\beta} u_{\alpha\beta} \ ,
\label{basic}
\end{eqnarray}
where $\B A$ is the usual elastic tensor, and $\mathrm{d} S$ is the area element on the boundary.  
The strain is related to the displacement field via 
\begin{equation}
 u_{\alpha\beta} = \frac{1}{2}\left(\partial_\alpha  d_\beta  + \partial_\beta d_\alpha \right) \ .
 \label{defu}
\end{equation} 

The total energy is minimized with respect to $\mathbf{d}$: 
\begin{equation}
\begin{split}
\delta_d F&= \delta_d \int \Lag \dif^3 x - \oint t^\beta \delta d_\beta \mathrm{d} S\\&=\int \dif^3 x  \sigma^{\alpha\beta}\delta u_{\alpha\beta}- \oint t^\beta \delta d_\beta \mathrm{d} S\ .
\end{split}
\label{StrainVar1}
\end{equation} 
\begin{figure}
	\centering
	\includegraphics[width=\linewidth]{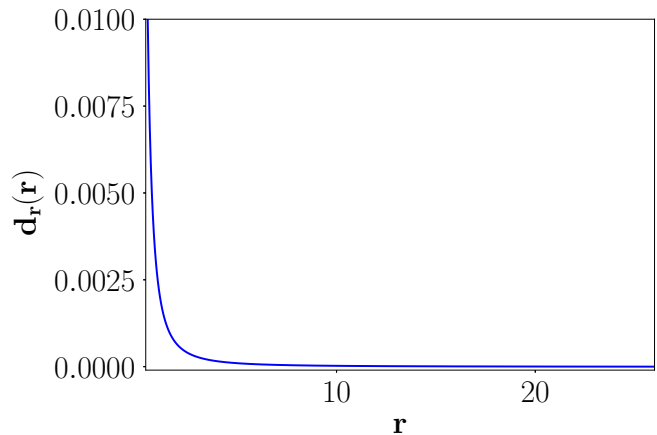}
	\caption{The analytic solution Eq.~(\ref{renelas}) for the radial displacement field that results from inflating an inner boundary $r_{\rm in}=2.17$ centered at the origin of a sphere with $r_{\rm out}=25.28$. Here $d_0=0.00084$.}
	\label{simple}
\end{figure}

Substituting Eq.~(\ref{defu}) in Eq.(\ref{StrainVar1}) and integrating by parts we get
\begin{equation}
\begin{split}
\delta_d F &= \int \dif^3 x \sigma^{\alpha\beta} \partial_\alpha \delta d_\beta- \oint t^\beta \delta d_\beta \mathrm{d} S\\ &= \oint \left(\sigma^{\alpha\beta} \, n_\alpha  - t^\beta\right) d_\beta \dif S -  \int \dif^3 x \partial_\alpha \sigma^{\alpha\beta} \delta d_\beta 
\end{split}
\end{equation} 
where $\B n$ is the unit vector in normal to the boundary. Hence
\begin{eqnarray}
\partial_\alpha \sigma^{\alpha\beta} &=& 0\nonumber\\
\sigma^{\alpha\beta} \, n_\beta \lvert_{\partial} &=& t^{\alpha} \ .
\label{equi}
\end{eqnarray}
Upon substituting the expression for the stress tensor in terms of strain, and then using the relation of strain and displacement, we find the equation for the displacement field in the form 
\begin{equation}
\Delta \mathbf{d} + \xi  \nabla  \left(\nabla \cdot \mathbf{d}\right)= 0 \ , \quad \xi  \equiv \frac{\lambda+\mu}{\mu} \ ,
\label{eq:EquilibriumDisp}
\end{equation}
where $\mu$ and $\lambda$ are the shear and bulk Lam\'e moduli. 

\subsection{Spherical inflation}
As explained in the introduction, we consider here a small sphere of radius $\rin$ at the center of a large sphere of radius $\rout$, with an imposed displacement $\mathbf{d}(\rin) = d_0 \hat{r}$ and $\mathbf{d}(\rout) = 0$. The spherical symmetry of implies that $\mathbf{d}(r) =d_r(r) \hat{r}$, in which case the equilibrium equation reduces to 
\begin{equation}
	\Delta {\mathbf{d}}=0 \ .
\end{equation}
The solution to this differential equation is
\begin{equation}
	d_r (r) = d_0 \frac{r_{\text{in}}^2 \left(r^3-r_{\text{out}}^3\right)}{r^2 \left(r_{\text{in}}^3-r_{\text{out}}^3\right)}\ .
	\label{renelas}
\end{equation}
In the bulk of the system, when $\rin< r<\rout$, the solution decays as $1/r^2$, as expected in standard elasticity theory. The graphic representation of this solution is exhibited in Fig.~\ref{simple}, 
showing that the solution goes like $1/r^2$ as is expected in standard elasticity theory.

\section{Geometric Charges in Three-Dimensions}
\label{charges}
In two-dimensional crystals, structural defects act as sources of stresses, and are realized as disclinations, dislocations and dislocation pairs \cite{80K}. 
As sources of stresses, defects act as elastic charge multipoles \cite{15MSK}. This observation allows a generalization of the notion of elastic charges to materials with no underlying structure, i.e. continuous and amorphous solids \cite{15MLAKS, 15KMS}.  The theory of \emph{geometric charges} provides a model-free definition for sources of stress as singularities of a curvature field describing  a reference geometry. This approach provides a full classification and multipole expansion of charges in arbitrary elastic-like solids. 
Within this framework, disclinations, dislocations and dislocation-pairs are nothing but monopolar, dipolare, and quadrupolar charge-multipoles of the reference curvature \cite{15MSK}. A direct result of the theory is a topological conservation law with the monopole and dipole being conserved \cite{15KMS}.  A physical consequence of this conservation law is that local deformation of a solid in a bounded region cannot induce net monopole or dipole charge, and therefore materials that respond by local plastic deformations will generically induce local quadrupolar charges. 

The key-point relevant to the current work is that the first non-conserved charge in amorphous solids is induced by local inelastic strain. While an extension of the theory of geometric charges does not exist in three-dimensions, we hypothesize that the first non-conserved point sources of stresses in three-dimensions is also induced by local inealstic strain, and therefore quadrupolar. This hypothesis is supported by observations in amorphous solids where localized plastic events practically form Eshelby-inclusions, that are induced by local strain  \cite{54Esh}.
Upon denoting the inealstic strain distribution, that is the quadrupolar charges, by $Q_{\alpha\beta} (\mathbf{x})$, a multipole expansion suggests that the divergence of quadrupoles field acts effectively as a dipole field. This is analogous to the divergence of polarization in dielectric media, which acts as a monopole field \cite{landau2013electrodynamics}. In three-dimensional solids, we define the dipole field
\begin{eqnarray}
	P_\alpha \equiv \partial^\beta Q_{\alpha\beta} \;.
\end{eqnarray}
The continuum description of quadrupolar charges as distributed inealstic strains, and the emergent dipole field as its divergence, are sufficient for developing the quadrupole and dipole screening theories in three-dimensional amorphous solids as presented in subsequent sections.

\section{Anomalous elasticity}
\label{dilquad}

In this section, we detail the derivation of equations for the displacement field in the presence of quadrupoles. As in the two-dimensional case, we will find either elasticity theory with renormalized elastic moduli or anomalous elasticity with qualitatively new behavior. We refer to the first ``quasi-elastic" since this theory is also emergent, having moduli that are dressed self-consistently by the response of the system. 

The starting point is an expression for the Lagrangian of the system, consistent with the underlying symmetries, presented up to quadratic order in the relevant fields. We note that plastic deformations in 3-dimensional solids can be described by their corresponding (quadrupolar) eigen-strain $Q^{\alpha\beta}$ \cite{54Esh}. 
The elastic energy stored in the system stems from three main contributions \cite{13DHP}. First, the energetic cost associated with the bare imposed stress field $U_\text{el}$. Second is the interaction of the induced quadrupoles with the elastic background $U_\text{Q-el}$. Lastly, there is the self-interaction of the quadrupoles, reflecting their nucleation cost. Explicitly 
\begin{equation}
\begin{split}
U = U_\text{el} + U_\text{Q-el} + U_\text{QQ} \ .
\end{split}
\label{eq:Udecomp}
\end{equation} 
where
\begin{equation}
\begin{split}
	U_\text{el} &=\int \dif^3 x \frac{1}{2} A^{\alpha\beta\gamma\delta} u_{\alpha\beta}u_{\gamma\delta}\\
	U_\text{Q-el} &=  \int \dif^3 x \Gamma^{\alpha\beta}_{\gamma\delta} u_{\alpha\beta}Q^{\gamma\delta} \\
		U_\text{QQ} &=\int \dif^3 x \frac{1}{2}\mathcal{F} \left(Q^{\alpha\beta},\partial_\beta Q^{\alpha\beta}\cdots\right) 
\end{split}
\label{eq:energydecomp}
\end{equation}
where  $\B \Gamma$ is an appropriate coupling tensor.  Here $\mathcal{F}$ represents the energy cost of the induced plastic quadrupoles, including their first and second gradient terms. As explained,
and in analogy with the two-dimensional case,
only the divergence of the quadrupole field contributes
to the charge density, and its curl is irrelevant here. Therefore below we employ $\mathcal{F} \left(Q^{\alpha\beta},\partial_\beta Q^{\alpha\beta}\right) $.
\subsection{Low quadrupoles density}
In the dilute quadrupoles limit, corresponding to large
energetic cost for nucleating dipoles, quadrupoles vary
slowly in space to avoid effective dipoles, hence
 $\C F=\mathcal{F} \left(Q^{\alpha\beta}\right) $. A Taylor expansion implies
 \begin{equation}
 	\mathcal{F} \left(Q^{\alpha\beta}\right) = \frac{1}{2}\Lambda_{\alpha\beta\gamma\delta} Q^{\alpha\beta}Q^{\gamma\delta}	
 \end{equation}
 Upon minimizing \eqref{eq:Udecomp}, with respect to the fundamental fields $d$ and $Q$, using  \eqref{eq:energydecomp}, we find 
\begin{equation}
	\begin{split}
		\delta_Q U= \delta_Q \int \Lag \dif^3 x=  \int \dif^3 x \left( \Lambda_{\alpha\beta\gamma\delta} Q^{\alpha\beta}+ \Gamma^{\alpha\beta}_{\gamma\delta} u_{\alpha\beta}\right)\delta Q^{\gamma\delta}\\
		\delta_d U= \delta_d \int \Lag \dif^3 x=  \int \dif^3 x \left( \sigma^{\alpha\beta}\delta u_{\alpha\beta} +   \Gamma^{\alpha\beta}_{\gamma\delta} Q^{\gamma\delta} \delta u_{\alpha\beta}\right)  
	\end{split}
	\label{StrainVar}
\end{equation}

From the first equation we get a linear screening relation (analogous to the linear relation between electric field and induced polarization in dielectric materials \cite{49Fro})
\begin{equation}
	Q^{\alpha\beta} = - \Lambda^{\alpha\beta\mu\nu}  \Gamma_{\mu\nu}^{\gamma\delta} u_{\gamma\delta} \equiv  - \tilde{\Lambda}^{\alpha\beta\gamma\delta} u_{\gamma\delta} \ ,
	\label{eq:Screeningrelation}
\end{equation}
where $\Lambda^{\alpha\beta\mu\nu}$ is the inverse of $\Lambda_{\alpha\beta\mu\nu}$. Note that this result indicates that the quadrupole field is not some mysterious invention, but an explicit 
response to the strain field.

Substituting in Eq.(\ref{StrainVar}) and integrating by parts we get
\begin{equation}
	\begin{split}
		\delta_d U &=  \int \dif^3 x \left( \sigma^{\alpha\beta} \partial_\alpha \delta d_\beta +   \Gamma^{\alpha\beta}_{\gamma\delta} Q^{\gamma\delta}\partial_\alpha \delta d_\beta\right) \\ &= \oint  \left(\sigma^{\alpha\beta} + \Gamma^{\alpha\beta}_{\gamma\delta} Q^{\gamma\delta}  \right) n_\alpha \delta d_\beta \dif S\\& -  \int \dif^3 x \partial_\alpha  \left(\sigma^{\alpha\beta} + \Gamma^{\alpha\beta}_{\gamma\delta} Q^{\gamma\delta}  \right)  \delta d_\beta \;.
	\end{split}
\end{equation} 
We note that the boundary conditions constrain an effective (dressed) stress field $\tilde{\sigma}^{\alpha\beta} = \sigma^{\alpha\beta} + \Gamma^{\alpha\beta}_{\gamma\delta} Q^{\gamma\delta}$ for which the equilibrium equation is 
\begin{equation}
	\partial_\alpha  \tilde{\sigma}^{\alpha\beta} = 0 \ ,
	\label{eq:Equilibrium}
\end{equation}
The effective elastic properties are obtained by substituting the constitutive relation (\ref{eq:Screeningrelation}) in the energy density (\ref{eq:Udecomp}):
\begin{equation}
	\begin{split}
		\Lag &=  \frac{1}{2} A^{\mu\nu\rho\sigma} u_{\mu\nu}u_{\rho\sigma} + 
		 \frac{1}{2} \Lambda_{\alpha\beta\gamma\delta} Q^{\alpha\beta}Q^{\gamma\delta}
		 + \Gamma^{\alpha\beta}_{\gamma\delta} u_{\alpha\beta}Q^{\gamma\delta} \\&=  \frac{1}{2} A^{\mu\nu\rho\sigma} u_{\mu\nu}u_{\rho\sigma} + 
		 \frac{1}{2} \Lambda_{\alpha\beta\gamma\delta} \left(- \tilde{\Lambda}^{\alpha\beta\mu\nu} u_{\mu\nu}\right) \left(- \tilde{\Lambda}^{\gamma\delta\rho\sigma} u_{\rho\sigma}\right)
		  \\&+ \Gamma^{\mu\nu}_{\gamma\delta} u_{\mu\nu}\left(- \tilde{\Lambda}^{\gamma\delta\rho\sigma} u_{\rho\sigma}\right) \equiv \frac{1}{2} \tilde{A}^{\mu\nu\rho\sigma}u_{\mu\nu}u_{\rho\sigma}
	\end{split}
\end{equation} 
with 
\begin{eqnarray}
	\tilde{A}^{\mu\nu\rho\sigma} = {A}^{\mu\nu\rho\sigma} +  \Lambda_{\alpha\beta\gamma\delta}  \tilde{\Lambda}^{\alpha\beta\mu\nu}  \tilde{\Lambda}^{\gamma\delta\rho\sigma} - 2 \Gamma^{\mu\nu}_{\gamma\delta} \tilde{\Lambda}^{\gamma\delta\rho\sigma}
\end{eqnarray}
we see that the re-normalization of the quadrupole-quadrupole interactions results with a linear constitutive relation between inducing stress and induced quadrupoles which then renormalizes the elastic tensor \cite{20NWRZBC}. This is the analog of the situation in dielectrics, where the dielectric constant is renormalized
by the induced dipoles \cite{49Fro}. In the next subsection we provide examples of this situation.
\begin{figure}
	\includegraphics[width=0.9\linewidth]{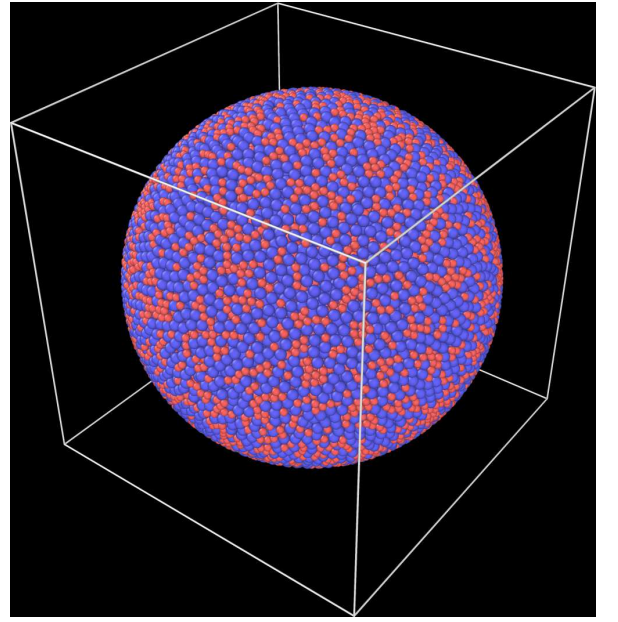}
	\caption{Typical configuration of our system of $N=42876$ bi-dispersed little spheres with volume fraction $\phi=0.647$.}
	\label{config}
\end{figure}

\subsection{Numerical Examples} 
\label{exam1}

\begin{figure}
	\includegraphics[width=1.0\linewidth]{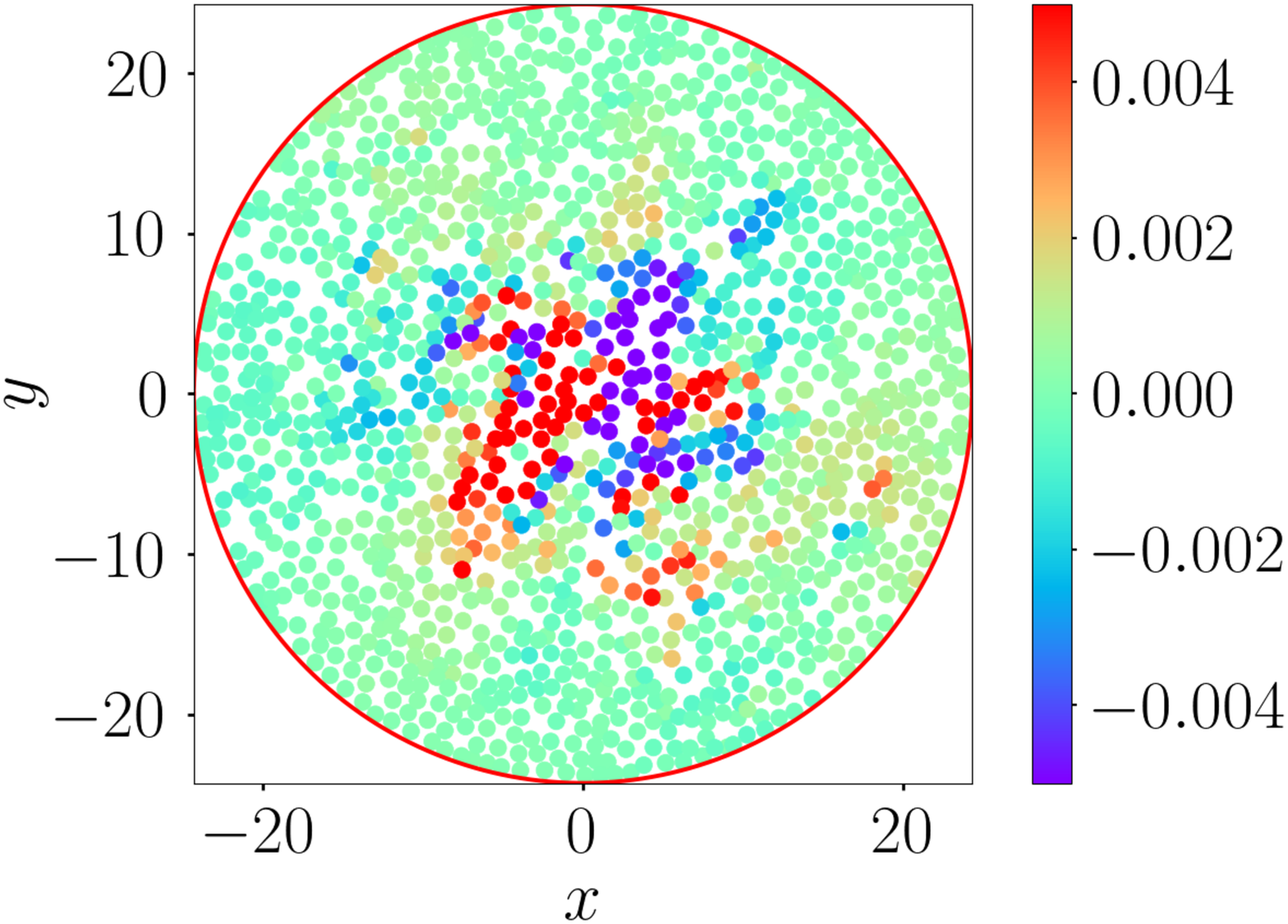}
	\includegraphics[width=0.9\linewidth]{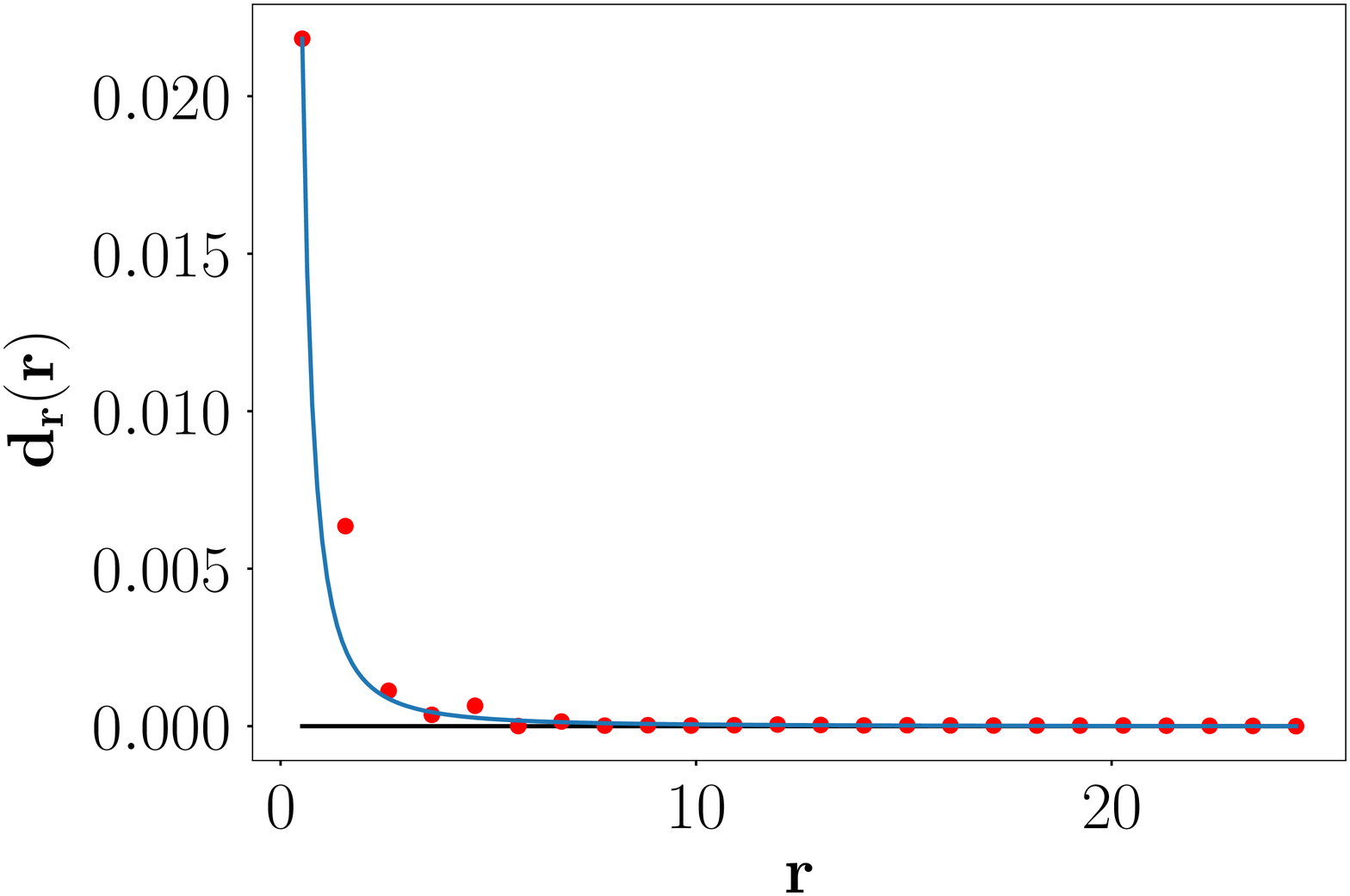}
	\caption{Displacement due to inflation of 30\%, volume fraction used here is $\phi=0.7$. Panel a: the displacement field shown on the $x-y$ plane at $z=0$. The color code represents the magnitude of the radial displacement field which is centralized around the origin, decaying towards the outer boundary. Here $r_{\rm in} = 0.72$,
		$r_{\rm out}= 24.29$ and 
		$d_0 = 0.0789$. Panel b: Spherically averaged displacement field.  Shown also is the fit of Eq.~(\ref{renelas}) to the angle-averaged displacement. }
	\label{qelastic}
\end{figure}

To test our theory of the effect of low density quadrupoles we employ the system described in Appendix \ref{proto}. A typical configuration of $N=42876$ bi-dispersed little spheres with volume fraction $\phi=0.647$ is shown in Fig.~\ref{config}.  The little spheres are interacting via a Hertzian normal force, and we choose the force constant large, $K_n=2\times 10^5$. This choice creates a stiff system that responses
to the inflation of $r_{\rm in}$ with a paucity of plastic events. Indeed this is what we see, cf. the displacement maps in panel (a) of Fig.~\ref{qelastic}.  Here the color code is the magnitude of the radial component of the displacement field. One can observe immediately that the displacement field is concentrated near
the inflated sphere and decays towards the outer boundary, as is expected in (renormalized) elasticity.

The spherical-averaged radial component of the displacement field $d_r(r)$ is computed from the data, and is shown together with the solution Eq.~(\ref{renelas}) in panel (b) of Fig.~\ref{qelastic} as a function of $r$. The spherical averages are evaluated on spherical shells, and the continuum approximation fits extremely well the measured data.

\section{High density of quadrupoles}
\label{highdensity}
At high quadrupole densities one cannot neglect the gradient terms. Upon denoting the elastic tensor dressed by the induced quadrupolar terms (as discussed in the last subsection) by $\tilde{\A}$,  the Lagrangian reads
\begin{equation}
	\begin{split}
		\Lag &=  \frac{1}{2} \tilde{\A}^{\mu\nu\rho\sigma} u_{\mu\nu}u_{\rho\sigma} + 
		\frac{1}{2} \Lambda_{\alpha\beta} \partial_\mu Q^{\mu\alpha}  \partial_\nu Q^{\nu\beta}
		+ \Gamma_{\alpha}^{\,\,\beta} \partial_\mu Q^{\mu\alpha} d_{\beta} \ .
	\end{split}
\end{equation} 
the last term here results from $U_{Q-el}$ in \eqref{eq:energydecomp} by integration by parts, using Eq.~(\ref{defu}).

Denoting the gradients on the quadrupoles as effectively induced dipoles  $P^\alpha \equiv \partial_\beta Q^{\alpha\beta}$, and minimizing with respect to the fundamental fields $d$ and $Q$ we find 
\begin{equation}
	\begin{split}
	\delta_Q \int \Lag \dif^3 x=  \int \dif^3 x \left( \Lambda_{\alpha\beta} P^{\alpha}+ \Gamma^{\alpha}_{\beta} d_{\alpha}\right)\delta P^{\beta}\\
	\delta_d \int \Lag \dif^3 x=  \int \dif^3 x \left( \sigma^{\alpha\beta}\delta u_{\alpha\beta} +   \Gamma^{\alpha}_{\beta} P^{\beta} \delta d_{\alpha}\right)  
	\end{split}
	\label{eq:StrainVar2}
\end{equation}

From the first equation we get a linear screening relation 
\begin{equation}
	P^{\alpha} = - \Lambda^{\alpha\beta}  \Gamma_{\beta}^{\gamma} d_{\gamma} \ .
	\label{eq:Screeningrelation2}
\end{equation}
Again we learn that the dipole field is not a mysterious construct, but rather a response to the displacement field. A direct measurement of the dipoles was presented in the context of two-dimensional systems in Ref.~\cite{22BMP}.

In the second equation we use again Eq.~(\ref{defu}) and integrate by parts. We get
\begin{equation}
	\begin{split}
		\delta_d U &=   \int \dif^3 x \left( \sigma^{\alpha\beta} \partial_\alpha \delta d_\beta    +   \Gamma_{\alpha}^{\beta} P^{\alpha} \delta d_{\beta}\right) \\ &= \oint  \sigma^{\alpha\beta} n_\alpha \delta d_\beta \dif S \\&-  \int \dif^3 x   \left(\partial_\alpha \sigma^{\alpha\beta} + \Gamma_{\alpha}^{\beta} P^{\alpha} \right)  \delta d_\beta 
	\end{split}
\end{equation} 
hence
\begin{equation}
	\partial_\alpha \sigma^{\alpha\beta} =- \Gamma_{\alpha}^{\beta} P^{\alpha}
	\label{eq:Equilibrium2}
\end{equation}
Combining \eqref{eq:Screeningrelation2} with \eqref{eq:Equilibrium2} we find
\begin{equation}
	\partial_\alpha \sigma^{\alpha\beta} =- \Gamma_{\alpha}^{\beta} P^{\alpha} = \Gamma_{\alpha}^{\beta}  \Lambda^{\alpha\mu}  \Gamma_{\mu}^{\gamma} d_{\gamma}
	\label{eq:ScreeningEquilibrium2}
\end{equation}
We see that the displacement field acts as a screening source in the equilibrium equation. We  now rewrite this equation by substituting the stress in terms of strain, and the strain in terms of the displacement. In isotropic and homogeneous materials the coupling tensors have the following forms
\begin{eqnarray}
	\begin{split}
		\A^{\alpha\beta\gamma\delta} &= \lambda  g^{\alpha\beta} g^{\gamma\delta} + \mu \left(g^{\alpha\gamma} g^{\beta\delta}  + g^{\alpha\delta} g^{\beta\gamma} \right)\\
		\Gamma_{\alpha}^{\beta} &= \mu_1 g^\alpha_\beta\\
		\Lambda^{\alpha\beta} &= \mu_2 g^{\alpha\beta}
	\end{split}
\end{eqnarray}
with $\B g$ the euclidean metric tensor, and $\mu_1$, $\mu_2$ being scalar coefficients.
Direct substitution yields
\begin{equation}
	\mu  \Delta \mathbf{d} + \left(\lambda + \mu \right) \nabla \left(\nabla\cdot \mathbf{d}\right) = \mu_1 \mathbf{P} = -\frac{\mu_1^2}{\mu_2} \mathbf{d}
\end{equation}
or in a simpler form
\begin{equation}
	\Delta \mathbf{d} + \left(1+\frac{\lambda}{\mu}\right) \nabla \left(\nabla\cdot \mathbf{d}\right) = -\frac{\mu_1^2}{\mu_2 \mu } \mathbf{d}
	\label{eq:DisplacementEquation}
\end{equation}
The screening effect is negligible when $\frac{\mu_1^2}{\mu_2 \mu } \ll 1$. 
Unlike the quadrupole screening, dipole screening leads to a qualitatively new equation. The appearance of the displacement field $\B d$ without gradients represents a breaking of translational symmetry. The parameter $\kappa$, defined via $\kappa^2\equiv \frac{\mu_1^2}{\mu_2 \mu }$, is an inverse scale. The appearance of a scale heralds the breakdown of classical elasticity theory, leading to qualitatively new mechanical responses as is shown next.

\subsection{Response to spherical inflation at the center}

\begin{figure}
	\includegraphics[width=1.0\linewidth]{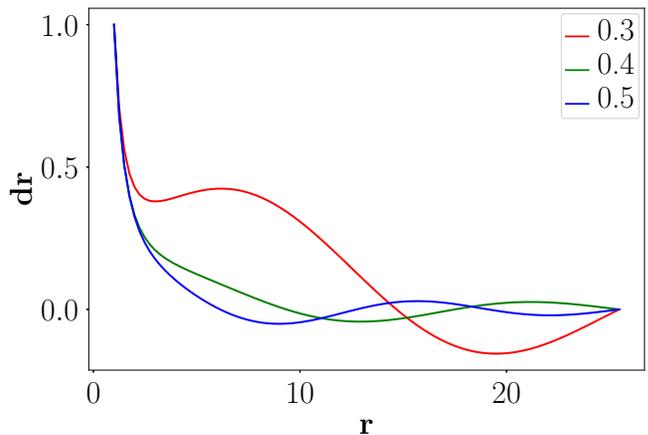}
	\caption{Solutions of the Bessel equation (\ref{Bessel}) for different values of the parameter $\kappa$. The radial component of displacement field for each $\kappa$ value is normalized to its maximum value at the inner boundary}
	\label{solution}
\end{figure}

Presently we discuss simulations that are identical in protocol and aim as those discussed in Subsect.~\ref{exam1}, but at much lower values of $K_n$, $K_n=2000$. A sphere which is closest to the origin was
inflated by a desired amount.  Upon assuming spherical symmetry Eq.~(\ref{eq:DisplacementEquation}) reduces to
\begin{equation}
	\Delta {\mathbf{d}} = -\frac{\mu_1^2}{\mu_2 \left(\lambda + 2\mu\right)} {\mathbf{d}} \equiv -\kappa^2 \, {\mathbf{d}}\ .
	\label{eq:InclusionEquation}
\end{equation}
In polar coordinates 
\begin{equation}
d_r'' + \frac{2 d_r'}{r}	-\frac{2 d_r}{r^2}=-\kappa ^2 d_r \ .
\label{Bessel}
\end{equation}
This is the Spherical Bessel equation. A solution 
of this equation satisfying $d_r(r_{\rm in})=d_0$, $d_r(r_{\rm out})=0$ reads
\begin{equation}
	d_r(r)  = d_0 \frac{ Y_1(r \, \kappa ) J_1(r_\text{out} \kappa )-J_1(r \, \kappa ) Y_1(r_\text{out} \kappa )}{Y_1( r_\text{in} \kappa ) J_1(r_\text{out} \kappa )-J_1(r_\text{in} \kappa ) Y_1(r_\text{out} \kappa )} \ .
	\label{amazing}
\end{equation}
Here $J_1$ and $Y_1$ are the Spherical Bessel functions of the first and second kind respectively. The solution of this equation for some different values of $\kappa$ are shown in Fig.~\ref{solution}. We note that the length scale that is apparent from the appearance of a maximum in Fig.~\ref{solution} is determined by the
inverse of $\kappa$. Whether a maximum will appear, or whether several maxima will show up, depends
on the ratio between $r_\text{out}$ and $\kappa^{-1}$. 

\subsection{Comparison with the simulations}

\begin{figure}
	\includegraphics[width=1.0\linewidth]{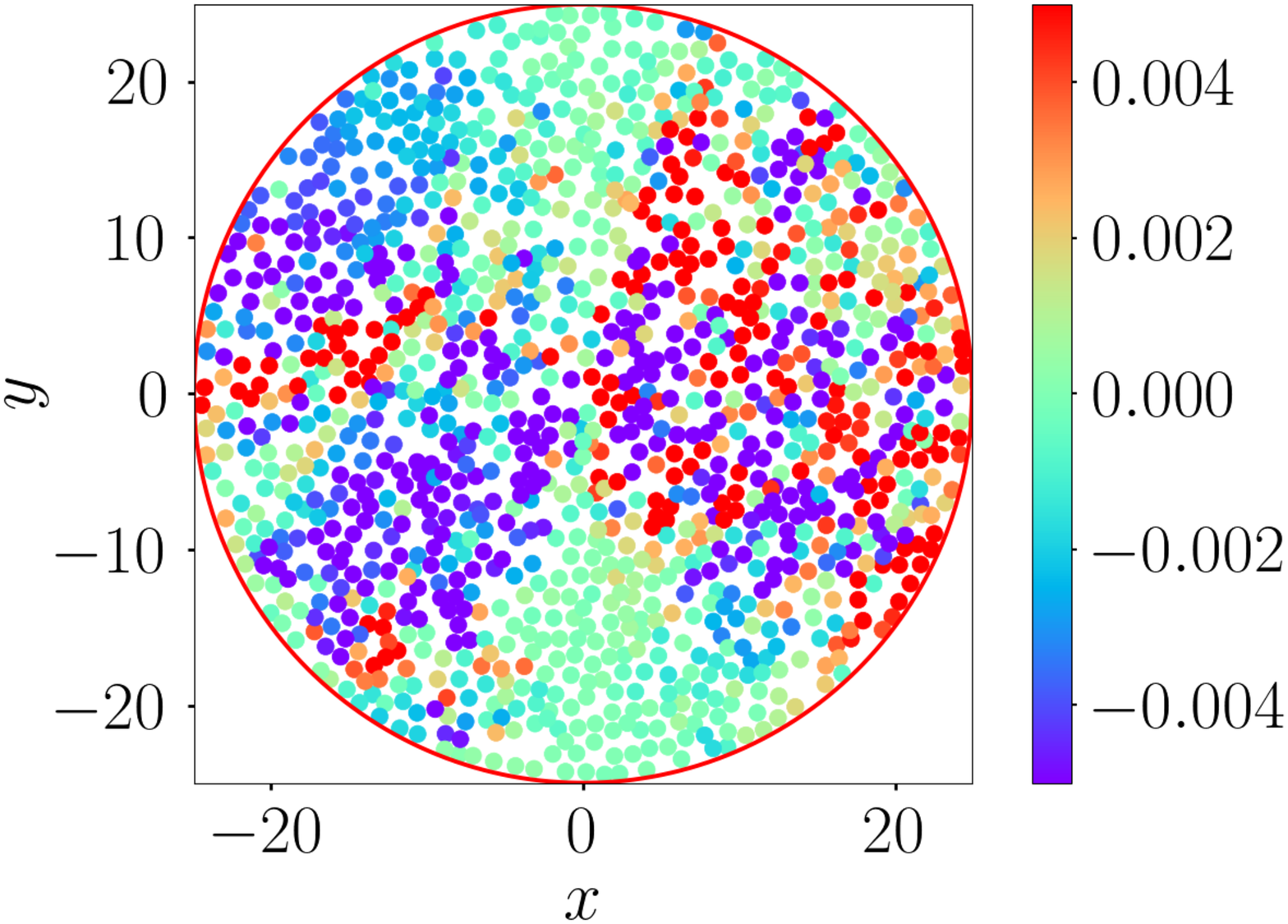}
		\includegraphics[width=1.0\linewidth]{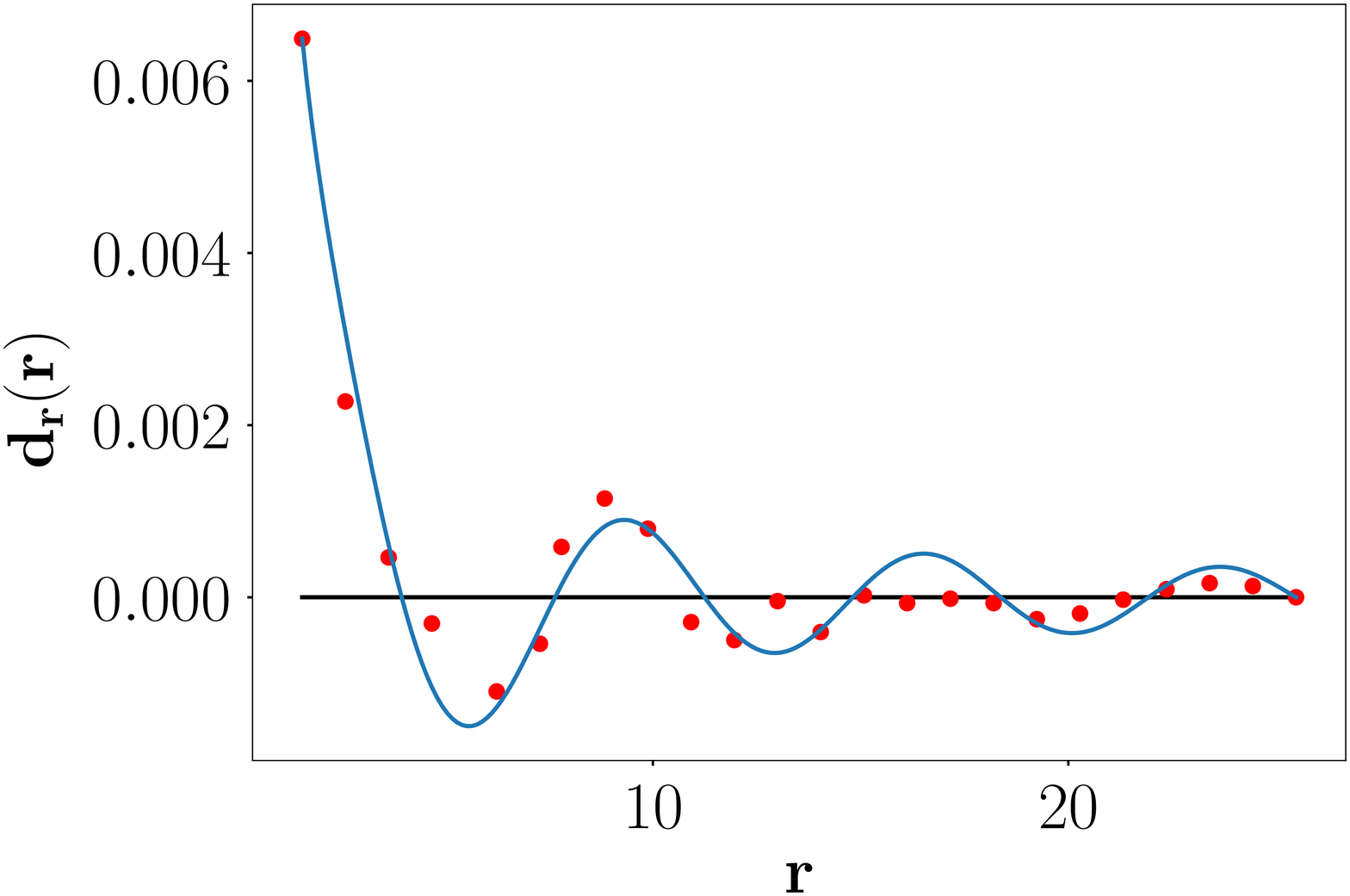}
	\caption{Radial displacement field with 10\% inflation, $\phi=0.647$. Panel a: Radial displacement field in a planar cross section of the three-dimensional sphere at $z=0$. One can see the qualitative difference from the map of Fig.~\ref{qelastic}, with activity going all the way to the outer boundary and displacement  pointing inward at places. Panel b: Comparison of the spherical averaged displacement field with $K_n=2000$ to the theory Eq.~(\ref{amazing}), using $\kappa=0.887$. Here  $\rin=1.56$ and $\rout=24.94$, with $d_0=0.0065$. }
	\label{anomalous}
\end{figure}

\begin{figure}
	\includegraphics[width=1.0\linewidth]{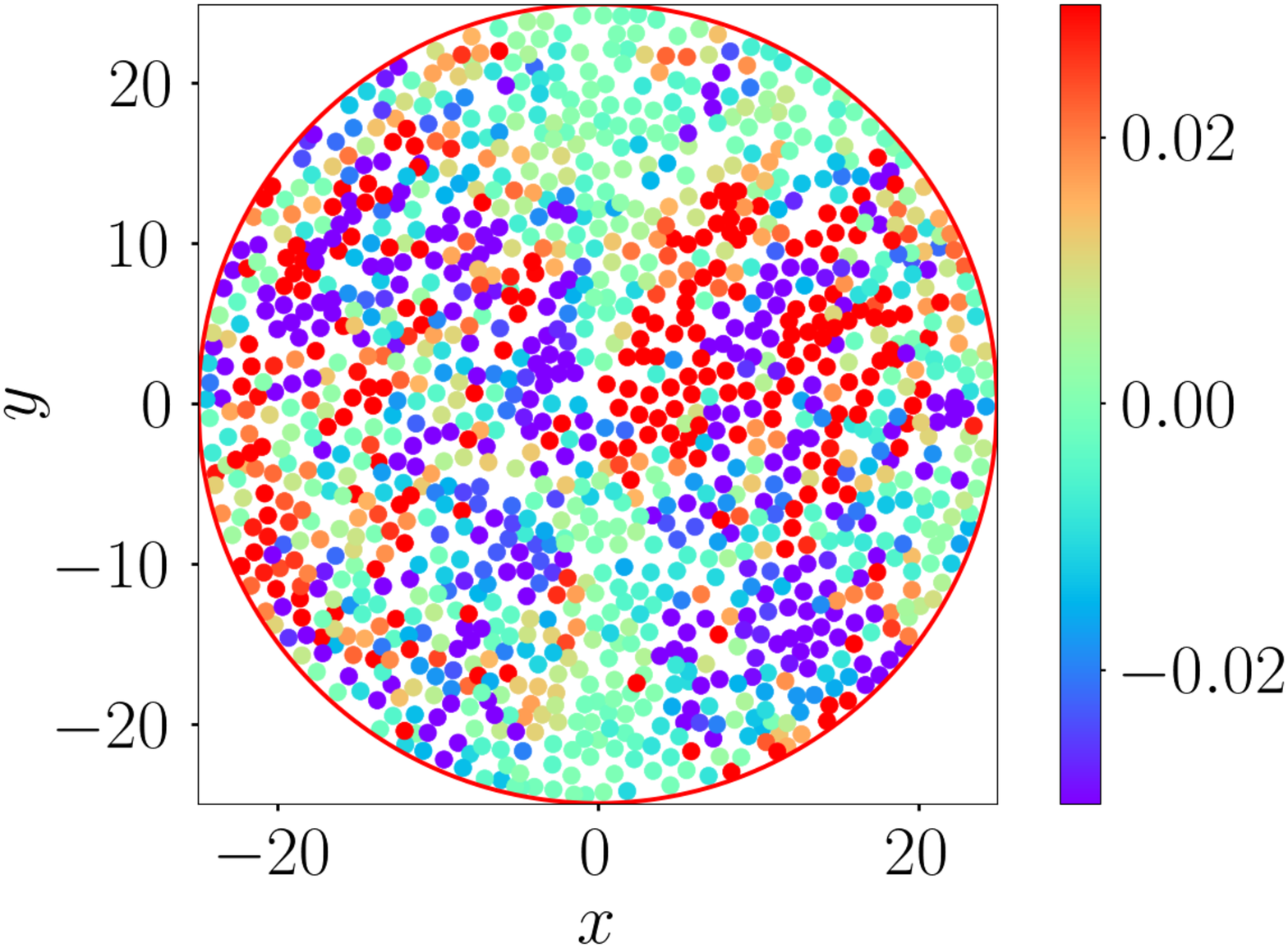}
	\includegraphics[width=1.0\linewidth]{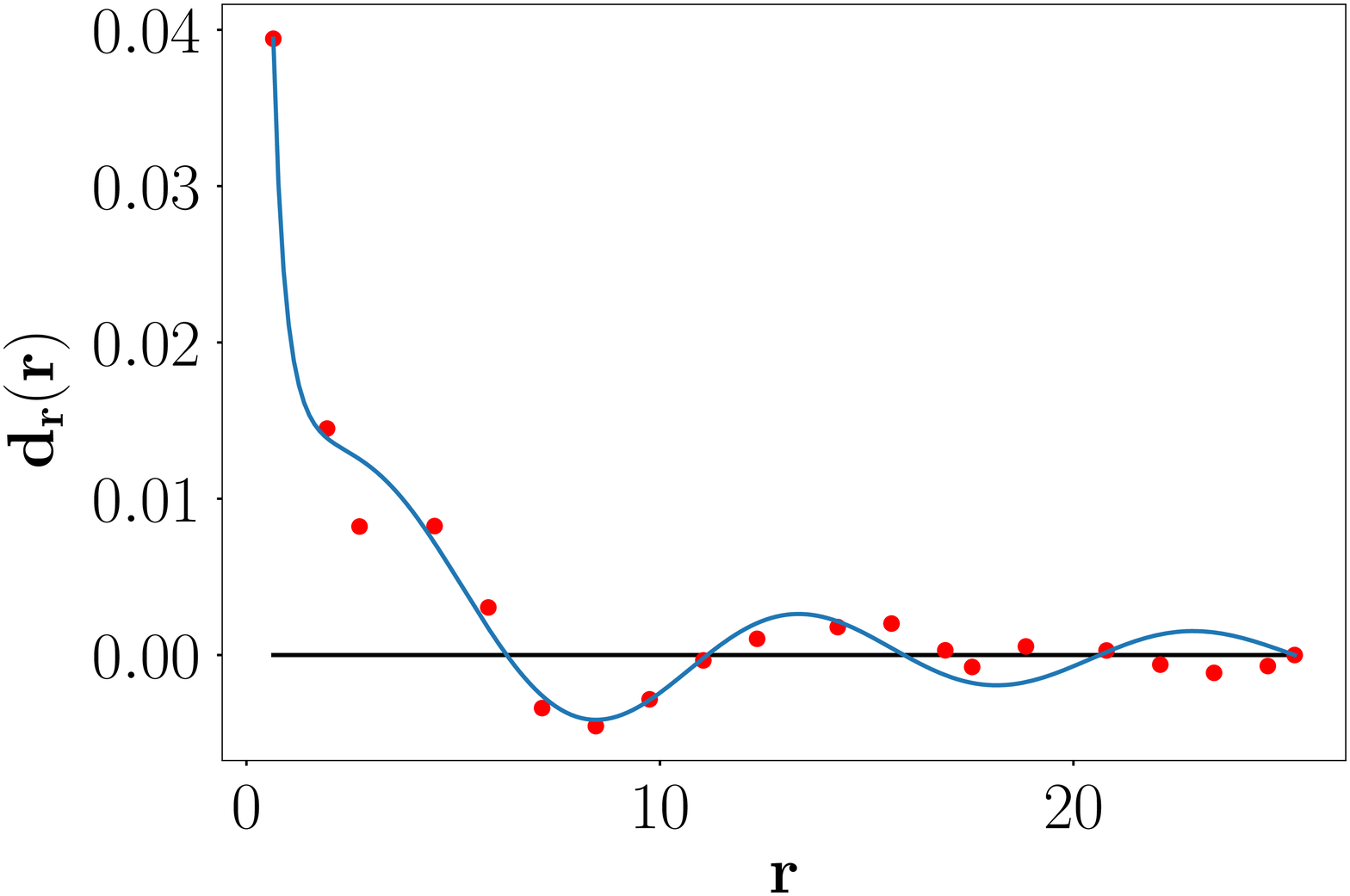}
	\caption{Radial displacement field with 30\% inflation, $\phi=0.649$. Panel a: Radial displacement 
	field in a planar cross section of the three-dimensional sphere at $z=0$. Panel b: Comparison of the 
	spherical averaged displacement field with $K_n=2000$ to the theory Eq.~(\ref{amazing}), using 
	$\kappa=0.669$. Here  $\rin=0.65$ and $\rout=25.35$, with $d_0=0.0394$ and $d_r(\rout)=0$. }
	\label{anomalous2}
\end{figure}

The simulations of our spherical system with configurations of binary little spheres serves
admirably to test the theory and the applicability of Eq.~(\ref{amazing}). Repeating the same
kind of simulations as before, but with a smaller packing fraction, $\phi=0.647$, we get, with inflation of 30\%, typical displacement fields as seen in panel (a) Fig.~\ref{anomalous}. Fitting
Eq.~(\ref{amazing}) to the spherical-averaged displacement field we get panel b of Fig.~\ref{anomalous}. The quality of the fit is more than satisfactory.
With packing fraction $\phi=0.649$ and inflation of 10\% we find the spherically averaged displacement field as seen in Fig.~\ref{anomalous2}.
Again, while not perfect, the fit is very satisfactory, especially taking into account
that we compare one single realization to a continuum theory!

\section{Summary and the road ahead}
\label{summary}

The upshot of this paper is that dipole screening is as important in three-dimensions as in two-dimensions, changing the equations of elasticity theory in a fundamental way. 
The ever existing plastic responses in three-dimensional amorphous solids should be carefully taken into account in assessing their implications for the mechanical properties of the host materials. The displacement field 
in response to a local source of stress can differ enormously from the prediction of classical
elasticity theory. In the examples shown above of inflation at the center, instead of observing the radial component of the displacement decaying like $1/r^{2}$, the displacement field can increase, decrease, and oscillate as a function of $r$. 

The response to inflation at the center can be either quasi-elastic or anomalous,  depending on the density of quadrupolar plastic
responses and its gradients (dipoles). In the present case we could go from one situation to the other by changing the force constant $K_n$ at a fixed volume ratio, or by changing the volume ratio for a fixed $K_n$. At large values of the force constant, at high pressure, the material is very stiff, with paucity of plastic responses. Then we find quasi-elastic response. At small values of $K_n$ or at small pressures there is a host of plastic events, and we then find anomalous responses,  in good agreement with the theory that is developed in this paper. 

An important issue that is left for future research is whether the transition from quasi-elastic to anomalous response is sharp or gradual,
as a function of material parameters or preparation protocol. From the data at hand it is still not possible to determine whether in three-dimensions the transition is  sharp, or a gradual cross over. This is worth further careful study, since this question is of considerable interest. The transition that we discuss is akin
to Kosterlitz-Thouless and Hexatic transitions, that share a change
from quadrupole dominated to dipole dominated regimes. It is known however that the latter transitions exist in two-dimensions only, and therefor if we
have a sharp transition in the present case it may constitute a first example of this kind of transition in three-dimensions. The research towards answering this question is under way, but it requires an enormous number of simulations, so the definite answer is left for a future publication. 

In fact, is worthwhile to reiterate that the kind of screening discussed here is analogous, but in fact richer, than the electrostatic counterpart. In electrostatics the players are electric charges and dipoles, whereas here we find dipoles
and quadrupoles. The screening of elasticity by quadrupoles only is analogous to dielectrics, having only a renormalization of the material parameters. The case of dipoles screening in elasticity, for which the displacement field changes qualitatively,
does not have an analog in electrostatics. Finally, monopoles may become relevant, but we did not deal with them explicitly since we expect that their appearance will be accompanied with
the melting (or un-jamming) of our amorphous solids.

In addition, there is the question of dynamics. We could in principle oscillate our inner sphere back and forth, inflating and deflating it dynamically. 
It would be very interesting and challenging to extend the theory presented above to take dynamics into account. We hope to be able to present such and extension in the near future. 

Finally, there is the issue of temperature. So far our studies are focused on athermal systems. The introduction of thermal fluctuations (and entropic effects)
may bring about new interesting aspects whose consequences will be picked up in future research. The addition of frictional forces
did not change the presence of quadrupolar and dipolar screening effects in 2-dimensions \cite{22MMPRSZ}, but it is worth while to examine whether
this is the case also in 3-dimensions.

\acknowledgments
This work had been supported in part by ISF under grant \#3492/21 (collaboration with China) and the Minerva Center for ``Aging, from physical materials to human tissues" at the Weizmann Institute. MM acknowledges support from the Israel Science Foundation (grant No. 1441/19).

\appendix

\section{System preparation and protocols}
\label{proto}

We investigate frictionless assemblies of small spheres that are at mechanical equilibrium,  prepared with a desired target pressure $P$ and confined in a spherical three-dimensional volume with a fixed outer spherical wall. Open source codes (LAMMPS \cite{95Pli}) are used to perform the simulations. Every simulation begins with a configuration of $N =42876 (35^3+1)$ bi-disperse disks placed randomly in a spherical volume with a radius, $\rout=25.48$ in L-J units.  Half of the small spheres have a radius $R_1 =0.5$ and the other half a radius $R_2=0.7$, both in L-J units.  One smaller disks is not placed randomly, but rather fixed to the center of the sphere. To reach a desired pressure we begin with a chosen packing fraction and the system is relaxed to mechanical equilibrium by solving Newton's second law of motion with damping. This process is carried out until the desired target pressure is reached and forces are minimized to values smaller than $10^{-7}$.

The normal contact force is Hertzian, following  the Discrete Element Method of Ref.~\cite{79CS}. The tangential contact force is zero as the system is frictionless. Let us consider two particles $i$ and $j$, at positions $\B r_i$, $\B r_j$ with velocities $\B v_i$, $\B v_j$. Two particles interact if and only if there is an overlap i.e. if the relative normal compression $\Delta_{ij}^{(n)}=D_{ij}-r_{ij}>0$, where $r_{ij}=|\vec r_{ij}|$, $\B r_{ij}=\vec r_i-\vec r_j$, $D_{ij}=R_i+R_j$, and {$R_i$, $R_j$} the radii of disks $i$ and $j$. The normal unit vector is denoted as $\vec n_{ij}=\vec r_{ij}/r_{ij}$. Normal component of the relative velocity at contact is given as,
\begin{equation}
	\begin{split}
		{\B v}^{(n)}_{ij}&= ({\B v}_{ij} .\B n_{ij})\,\B n_{ij}  
	\end{split}
\end{equation}
The normal force exerted by grain $j$ on $i$ is
\begin{equation}
	\begin{split}
		\B F^{(n)}_{ij}&=k_n\Delta_{ij}^{(n)}\B n_{ij}-\frac{\gamma_n}{2} \B {v}^{(n)}_{ij}
	\end{split}
\end{equation}
where
\begin{equation}
	\begin{split}
		k_n &= k_n'\sqrt{ \Delta^{(n)}_{ij} R^{-1}_{ij}} \ ,\\
		\gamma_{n} &= \gamma_{n}^{'}  \sqrt{ \Delta^{(n)}_{ij} R^{-1}_{ij}}\ ,
	\end{split}
\end{equation}
with $R_{ij}^{-1}\equiv R_i^{-1}+R_j^{-1}$. $k_n^{'}$ is the normal spring stiffness. The parameter $\gamma_n^{'}=500$ is the viscoelastic damping constant.  
In the current paper, the stiffness $k'_n$ varies from small to large to see the change from anomalous elasticity to quasi-elastic responses. The mass of each disk is $m=1$ in L-J units. 

After achieving a mechanically stable configurations at a target pressure, we inflated the disk at the center by a desired percentage varying between 10\% and 30\%.  We then measure the displacement field that is induced by this inflation. The data shown
throughout this paper is the spherical average of the radial component
of this measured displacement field.

\section{Microscopic derivation of Eqs.(\ref{eq:energydecomp})}
\label{micro}

The detailed representation of the energy was presented in Ref.~\cite{13DHP}.
We show now that the present equations are equivalent. As an example consider the first term
in \eqref{eq:energydecomp}:
\begin{equation}
	\begin{split}
		U_\text{el} &=\int \dif^3 x \frac{1}{2} A^{\alpha\beta\gamma\delta} u_{\alpha\beta}(x)u_{\gamma\delta}(x) \\&=\int \dif^3 x \frac{1}{2} A^{\alpha\beta\gamma\delta} u^\text{el}_{\alpha\beta}(x) u^\text{el}_{\gamma\delta}(x)   \\&+ \int \dif^3 x  A^{\alpha\beta\gamma\delta}  u^\text{el}_{\alpha\beta}(x) u^\text{Q}_{\gamma\delta}(x)  \\& + \int \dif^3 x \frac{1}{2} A^{\alpha\beta\gamma\delta}  u^\text{Q}_{\alpha\beta}(x) u^\text{Q}_{\gamma\delta}(x)
	\end{split}
	\label{eq:Ueldecomp}
\end{equation} 
Here the first term corresponds to the self energy associated with the bare elastic fields. The second term 
vanishes identically. To see this we have to substitute the explicit expression for $G^u_{\gamma\delta\mu\nu}$ and perform integration by parts twice with respect to $x$ to obtain an integrand proportional to $\varepsilon^{\alpha\mu}\varepsilon^{\beta\nu} \partial_{\mu\nu} u^\text{el}_{\alpha\beta}$, which is the compatibility condition on the bare strain.

The last term describes the interactions between induced quadrupoles located at different points:
	\begin{eqnarray}
			&&\int \dif^3 x \frac{1}{2} A^{\alpha\beta\gamma\delta}  u^\text{Q}_{\alpha\beta}(x) u^\text{Q}_{\gamma\delta}(x)\nonumber\\&&  =  \int \dif^3 x \dif^3 x' \dif^3 x''  \frac{1}{2} A^{\alpha\beta\gamma\delta}  G_{\alpha\beta\mu\nu}^u(x-x')\nonumber\\&&\times G_{\gamma\delta\rho\sigma}^u(x-x'') Q^{\mu\nu}(x') Q^{\rho\sigma}(x'') \nonumber\\&&=  \int \dif^3 x' \dif^3 x''  \frac{1}{2}  Q^{\mu\nu}(x') Q^{\rho\sigma}(x'') \nonumber \\&&\times \int \dif^3 x A^{\alpha\beta\gamma\delta}  G_{\alpha\beta\mu\nu}^u(x-x') G_{\gamma\delta\rho\sigma}^u(x-x'') \nonumber \\&&\equiv   \int \dif^3 x' \dif^3 x''  \frac{1}{2}  Q^{\mu\nu}(x') Q^{\rho\sigma}(x'') \Lambda_{\mu\nu\rho\sigma} (x'-x'')
		\label{eq:uQuQ}
	\end{eqnarray} 

This expression is ill defined in the case $x'-x''\to 0$. Therefore renormalization techniques are required, where a cutoff length scale is introduced to regularize the integral, representing the quadrupoles core size. This result with an additional term describing quadrupoles self interactions, as in the second expression in \eqref{eq:energydecomp}. 

The third term in \eqref{eq:energydecomp} has two contributions. A quadrupole-quadrupole interaction term, correcting the coefficient of  \eqref{eq:uQuQ}, and a quadrupole-strain term
\begin{equation}
	\begin{split}
		U_\text{Q-el} &=  \int \dif^3 x \Gamma^{\alpha\beta}_{\gamma\delta} u_{\alpha\beta}Q^{\gamma\delta} =  \int \dif^3 x \Gamma^{\alpha\beta}_{\gamma\delta} u^\text{el}_{\alpha\beta}Q^{\gamma\delta} \\&+  \int \dif^3 x \dif^3 x' \Gamma^{\alpha\beta}_{\gamma\delta}G^ u_{\alpha\beta \mu\nu}(x-x')  Q^{\gamma\delta}(x) Q^{\mu\nu}(x')
	\end{split}
\end{equation} 
Wrapping it all together we find 
\begin{equation}
	\begin{split}
		U &=\int \dif^3 x \frac{1}{2} A^{\alpha\beta\gamma\delta} u^\text{el}_{\alpha\beta}(x) u^\text{el}_{\gamma\delta}(x)   \\& + \int \dif^3 x' \dif^3 x''  \frac{1}{2}  Q^{\mu\nu}(x') Q^{\rho\sigma}(x'') \Lambda_{\mu\nu\rho\sigma} (x'-x'') 
		\\&+\int \dif^3 x \frac{1}{2} \Lambda_{\alpha\beta\gamma\delta} Q^{\alpha\beta}Q^{\gamma\delta} +\int \dif^3 x \Gamma^{\alpha\beta}_{\gamma\delta} u^\text{el}_{\alpha\beta}Q^{\gamma\delta}
	\end{split}
\end{equation} 
This is exactly the form of the energy functional in \cite{13DHP}, and therefore we focus on this functional form  in \eqref{eq:Udecomp},\eqref{eq:energydecomp}.

\bibliography{3dimensions}

\end{document}